\documentclass[12pt]{article}

\usepackage{scicite}

\usepackage{times}

\usepackage{amsmath}
\usepackage{amsfonts}
\usepackage{amssymb}
\usepackage{graphicx}

\usepackage{color}

\newenvironment{sciabstract}{%
\begin{quote} \bf}
{\end{quote}}

\title{Physical learning beyond the quasistatic limit}

\author
{Menachem Stern$^1$, Sam Dillavou$^1$, Marc Z. Miskin$^2$, \\Douglas J. Durian$^1$ and Andrea J. Liu$^1$\\
\\
\normalsize{$^1$Department of Physics and Astronomy, University of Pennsylvania,}\\
\normalsize{Philadelphia, PA 19104}\\
\normalsize{$^2$Department of Electrical and Systems Engineering, University of Pennsylvania,}\\
\normalsize{Philadelphia, PA 19104}\\
}

\date{}

\begin{document} 


\baselineskip24pt


\maketitle



\begin{sciabstract}
Physical networks, such as biological neural networks, can learn desired functions without a central processor, using \emph{local} learning rules in space and time to learn in a fully distributed manner. Learning approaches such as equilibrium propagation, directed aging, and coupled learning similarly exploit local rules to accomplish learning in physical networks such as mechanical, flow, or electrical networks. In contrast to certain natural neural networks, however, such approaches have so far been restricted to the quasistatic limit, where they learn on time scales slow compared to their physical relaxation. This quasistatic constraint slows down learning, limiting the use of these methods as machine learning algorithms, and potentially restricting physical networks that could be used as learning platforms. Here we explore learning in an electrical resistor network that implements coupled learning, both in the lab and on the computer, at rates that range from slow to far above the quasistatic limit. We find that up to a critical threshold in the ratio of the learning rate to the physical rate of relaxation, learning speeds up without much change of behavior or error. Beyond the critical threshold, the error exhibits oscillatory dynamics but the networks still learn successfully.
\end{sciabstract}


\section*{Introduction}

There is a class of physically-realizable learning approaches that do not require a central processor and can be implemented, at least in principle, in real networks such as biological neuron, mechanical, flow or electrical resistor networks. In order for such physical networks to learn on their own, they cannot minimize an arbitrary cost function by gradient descent since that is a global process that requires knowing all the microscopic details at once, carrying out the global computation of gradient descent, and then manipulating networks at the microscopic (node or edge) level. Rather, approaches such as contrastive learning~\cite{movellan1991contrastive,lopez2021self,MARTIN2021102222,pashine2021local}, equilibrium propagation~\cite{scellier2017equilibrium,kendall2020training}, directed aging~\cite{pashine2019directed, hexner2019effect,hexner2019periodic, stern2018shaping,stern2020continual,stern2020supervised} and coupled learning~\cite{stern2021supervised,dillavou2021demonstration} use local rules, in which learning degrees of freedom (e.g. the conductances of edges in electrical networks of variable resistors) respond to physical degrees of freedom (e.g. node voltages) which automatically adjust themselves to minimize a physical cost function (e.g. difference in dissipated power), automatically performing an approximation to gradient descent on the cost function. Until now, the use of local rules has been restricted to the limit in which the physical degrees of freedom equilibrate rapidly compared to the learning degrees of freedom. This quasistatic or near-quasistatic condition of slow learning has also limited the utility of local learning approaches such as equilibrium propagation as machine learning algorithms on the computer, by rendering them too slow to compete with global approaches such as backpropagation~\cite{bartunov2018assessing}. 

It has been shown recently that learning can occur beyond the strict quasistatic limit, and that continual updating of the learning degrees of freedom is possible as long as the learning rate is sufficiently low~\cite{ernoult2019updates,ernoult2020equilibrium,scellier2021deep}. However, certain biological networks are capable of learning at rates not just close to the quasistatic limit but far above it. For instance, some natural neural networks may reach a steady spiking state on a timescale that is similar to synaptic plasticity processes~\cite{zucker2002short,kanai2005perceptual,marom2010neural,sagi2012learning}. If a system must adapt to a rapidly fluctuating environment supplying the training signals, learning must occur on timescales similar to that of physical adjustment. This may be the case in the humoral immune response, where antibodies are formed rapidly after infection, even before the onset of symptoms~\cite{lemon1983serum,world2017guidelines,banatvala2004rubella}. 

In this work, we explore learning as a function of the scaled learning rate $\mathcal{R}$, the ratio of the rates of evolution of the learning and physical degrees of freedom. We specifically consider the task of allosteric response on electrical resistor networks that implement coupled learning, but our approach can be applied to other tasks on any physical learning network. We find that local learning rules can lead to effective learning not only in the quasistatic limit as previously assumed ($\mathcal{R} \ll 1$), but also for values of $\mathcal{R}$ near and even far above unity. We show experimentally and numerically that up to a critical threshold $\mathcal{R}_c \sim 1$, learning speeds up with essentially no change in the error or behavior. Beyond $\mathcal{R}_c$ the system still learns but at the cost of a higher error and oscillatory learning dynamics.  At still higher learning rates $\mathcal{R} \gg 1$, the achievable error drops once more, vanishing in the limit of infinitely fast learning $\mathcal{R} \rightarrow \infty$. It is notable that even for systems \emph{trained} far from equilibrium, the error in \emph{performing} the allosteric task in equilibrium is minimized as well.

We analyze the system theoretically to understand these results. 
Overall, our results suggest that learning in physical systems is achievable far from equilibrium, well above the quasistatic limit. This potentially allows for massive reduction in training times, relaxing limitations on local learning approaches as machine learning algorithms and opening up the use of coupled learning for training physical networks with slow physical dynamics.

\section*{Results}

\subsection*{Quasistatic and non-equilibrium coupled learning}

Recent learning methods in physical systems generally assume that the learning process is decoupled from physical processes. For example, the approach of directed aging for mechanical networks assumes that the physical degrees of freedom (the node positions) equilibrate to minimize the physical cost function (the elastic energy) on time scales that are fast compared to the evolution of the learning degrees of freedom (spring constants, equilibrium lengths, etc.).   

In such frameworks~\cite{scellier2017equilibrium,pashine2019directed,hexner2019effect,hexner2019periodic,kendall2020training,stern2021supervised,pashine2021local} the physical degrees of freedom $v_a$ on the nodes, indexed by $a$, first equilibrate to values $v_a^{*}$, and the learning degrees of freedom $G_j$ on the edges, indexed by $j$, are then modified according to $v_a^{*}$. The physical degrees of freedom then equilibrate again, spurring further change to the learning ones. In particular, both equilibrium propagation~\cite{scellier2021deep} and coupled learning~\cite{stern2021supervised} compare two equilibrium states (the free state $v_a^{*F}$ and the clamped state $v_a^{*C}$) to derive local learning rules that adjust the learning degrees of freedom. Likewise, approaches applied in the lab to mechanical networks~\cite{pashine2021local} and electrical resistor networks~\cite{dillavou2021demonstration} use learning rules comparing two different equilibrium states. In what follows, our notation is specific for electric resistor networks (with voltage values $v_a$ on nodes indexed by $a$ and conductances $G_j$ on edges indexed by $j$), but can easily be generalized to other types of physical networks~\cite{stern2021supervised}.

Throughout this paper, we will treat the case where the electrical network is to learn a desired ``allosteric" task in which a set of inputs $V_S$ are applied to designated source nodes, and the network responds by producing desired outputs $V_T$ at a designated set of target nodes. 
The free state is characterized by a vector $v_a^{*F}$, which is the set of all of the physical degrees of freedom, the voltages, on the nodes of the network when a set of input voltages $V_S$ are applied to the source nodes and the network equilibrates to produce the outputs voltage values $v_T^{*F}$ at the targets. Here the $^*$ notation indicates that the node voltages are equilibrated so that the physical cost function (dissipated power) is minimized; this is the quasistatic condition. The clamped state is defined by applying the inputs $V_S$ to the source nodes \emph{and} a prescribed set of input voltages $v_T^{C}$ at the target nodes. The values $v_T^{C}$ are chosen to be nudged slightly from $v_T^{*F}$ toward the desired output values so that $v_T^{C}=v_T^{*F} + \eta [V_T - v_T^{*F}]$, with $\eta \ll 1$. The coupled learning rule is given by
%
\begin{equation}
\begin{aligned}
\dot{G_j} = \gamma_{\ell} \eta^{-1} \frac{\partial}{\partial G_j} \{\mathcal{P}^{*F}(V_S,v_T^{*F})
- \mathcal{P}^{*C}(V_S,v_T^{C})\},
\end{aligned}
  \label{eq:GeneralRule}
\end{equation}
where the learning degrees of freedom are the edge conductances $G$, indexed by $j$,$\mathcal{P}^{*F}$ and $\mathcal{P}^{*C}$ are the power dissipated by the system in the equilibrium free and clamped states, respectively, and $\gamma_\ell$ is a learning rate. Note that since the power $\mathcal{P}$ can be written as a sum over edges, Eq.~\ref{eq:GeneralRule} is spatially local. An approximation to this learning rule has been implemented in the laboratory with an electronic circuit of variable resistors, which can be trained successfully to perform diverse tasks~\cite{dillavou2021demonstration}. The function $\mathcal{C}^*\equiv \eta^{-1}[\mathcal{P}^{*C}-\mathcal{P}^{*F}]$ is known as the \emph{contrastive} function, and has been shown to approximate the mean-squared-error cost function $C^*\sim (v_T^{*F} - V_T)^2$ in the limit $\eta\rightarrow 0$~\cite{scellier2017equilibrium,stern2021supervised}.

Suppose instead that the network learns (updates its edge conductances) 
while the node voltages (physical degrees of freedom) are still equilibrating. For simplicity we assume that the physical dynamics that relax the physical degrees of freedom are overdamped, with decay rate $\gamma_p$.  

\begin{equation}
\begin{aligned}
\dot{v}^{F}_a =& -\gamma_p \partial_{x} \mathcal{P}^F\\
\dot{G}_j =& -\gamma_\ell \partial_{g} \mathcal{C}.
\end{aligned}
  \label{eq:CoupledEquations}
\end{equation}

Note that for physical systems, computing such derivatives is linear in network size, so the computational efficiency of learning using Eq.~\ref{eq:CoupledEquations} is comparable to back-propagation on feed-forward neural networks. The scaled learning rate $\mathcal{R} \equiv \gamma_\ell/\gamma_p$ is the key quantity that we vary here. The limit $\mathcal{R} \rightarrow 0$ is the familiar quasistatic limit of coupled learning, in which the node voltages $v_a$ evolve infinitely rapidly compared to the edge conductances $G_j$. 

 The physical cost function, the power $\mathcal{P}^F$ in Eq.~\ref{eq:CoupledEquations} is the total power dissipated by the network, and can be written as a sum over edges: $\mathcal{P}=\sum_j G_j \Delta v_j^2/2$, where the voltage drop $\Delta v_j$ over each edge is defined by the difference in voltage between the two nodes connected by that edge $\Delta v_j\equiv \Delta_{ja} v_a$, where $\Delta_{ja}$ is the incidence matrix.
Eq.~\ref{eq:CoupledEquations} can therefore be simplified to

\begin{equation}
\begin{aligned}
\dot{v}_a^{F,C} =& -\gamma_p \sum_j \Delta_{aj}^T G_j \Delta_{ja} v_a^{F,C} \\
\dot{G}_j =& \gamma_\ell \eta^{-1} \{[\Delta v_j^{F}]^2 -  [\Delta v_j^{C}]^2\},
\end{aligned}
  \label{eq:CoupledFlow}
\end{equation}

Both Eq.~\ref{eq:GeneralRule} and Eq.~\ref{eq:CoupledFlow} are nontrivial to implement experimentally because it is impossible to apply simultaneously two different sets of boundary conditions (free and clamped). To circumvent this without storing information in any memory, the laboratory realization~\cite{dillavou2021demonstration} of quasistatic coupled learning (Eq.~\ref{eq:GeneralRule}) introduced two identical networks subjected to the different boundary conditions. In the nonequilibrium case, Eq.~\ref{eq:CoupledFlow} can likewise be implemented using two identical networks, a free network $v_a^F(t)$ and a clamped network $v_a^C(t)$, both evolving under the same overdamped physical dynamics and having the same edge conductances $G_j(t)$. As in Ref.~\cite{dillavou2021demonstration}, the difference between the two networks is the boundary conditions (applied external constraints): While both networks are constrained at the source(s) or input(s) by $v^{F,C}_S=V_S$, the clamped network is further constrained at the target(s) or output(s) by $v^C_T(t)=v^F_T(t)+\eta[V_T -v^F_T(t)]$.  

\subsection*{Experimental results}

To study nonequilibrium learning in the lab, we adapt the learning network introduced in Ref.~\cite{dillavou2021demonstration} that approximates quasistatic coupled learning~\cite{stern2021supervised}. Here we lift the quasistatic condition by slowing down relaxation of the node voltages, introducing capacitors in parallel to each variable resistor to realize the overdamped physical dynamics of Eq.~\ref{eq:CoupledFlow}. Edges of the network with attached capacitors are shown in Fig. 1(a). The charging of capacitors increases the time required to reach steady current in the network, but in other respects the experiment is almost\footnote{The update rule is slightly different, now consisting of the evaluation of $\mathrm{XOR}(\Delta v_j^F>\Delta v_j^C, \Delta v_j^F > - \Delta v_j^C)$ whereas before it was $\mathrm{XOR}(\Delta v_j^F>\Delta v_j^C, \Delta v_j^C>0)$. We have observed no effect from this change, but implemented it to adhere more closely to the original coupled learning rule.} as described in Ref.~\cite{dillavou2021demonstration}. The network is initialized by imposing the inputs and allowing the free network to reach equilibrium. The clamped boundary condition is applied by imposing both the desired input and output values and the clamped network is allowed to reach equilibrium as well. The training process then commences as in Ref.~\cite{dillavou2021demonstration}, but with a pause $1/\gamma_{\ell}$ prior to each update of the clamped voltages as they are adjusted towards the desired target values. (see Methods for more information).

The system's physical relaxation time changes during training as the edge resistances evolve. Regardless of these changes, the relaxation time will be proportional to the in-line capacitance $F$ added across every resistor. In each experiment $c$ takes one of four values, 2.2, 22, 220, or 2200 $\mu F$. We estimate a typical physical relaxation rate using resistance $R_0 \equiv 10K\Omega$ (each resistor starts at $50K\Omega$ and can be varied through the range $781\Omega-100K \Omega$). Thus, the two rates in the system are the learning rate $\gamma_{\ell}$ and the physical relaxation rate $\gamma_{p}=(R_0 c)^{-1}$ so that the scaled relaxation rate, or ratio of the rate of learning to that of physical relaxation, is $\mathcal{R}=R_0 c\gamma_{\ell}$.

We train the network to learn a two-target, two-source allosteric task (Fig. 1(b)), varying the capacitance $F$ and learning rate $\gamma_{\ell}$ to adjust $\mathcal{R}$. We define the instantaneous nonequilibrium error as the mean-squared difference between the free state outputs and the desired outputs $C(t)\equiv \sum_T(V_T-v_T^{F}(t))^2$, normalized by its initial value, as measured in real time so that the output voltages reflect the network's current nonequilibrium state. In the quasistatic regime, the scaled learning rate is low, $\mathcal{R} \ll 1$ and the system learns the task, as shown by the typical mean squared error (MSE) curves in red in Fig. 1(c) and (d), which eventually hover at the noise floor. In this regime, changing the ratio of relaxation rates, $\mathcal{R}$, does not affect the number of training steps required to learn the task $\tau$ (see collapse of the two reddest curves in (Fig. 1(c)) but changes the real time required to learn (see the same two curves in Fig. 1(d)). As $\mathcal{R}$ increases, the system's behavior becomes qualitatively different, at first taking more training steps to reach low error, and then entering into a regime with perpetual oscillations ($\mathcal{R} \gtrsim 1$), as seen best in the blue curve in Fig. 1(d). These oscillations widen the distribution of errors observed, as shown in Fig. 1(e), and increase the number of training steps required for the system to fall below an arbitrary normalized nonequilibrium error threshold of $C(\tau)=10^{-3}C(0)$ for $10$ training steps (this 10-step requirement yields more consistent results by eliminating the effect of `lucky' experimental runs in the fast-learning regime, in which the system flies through a very low-error state but fails to maintain it). In Fig. 1(f) we see two regimes: for $\mathcal{R} \ll 1$, the scaled learning time is well described by $\tau \sim \mathcal{R}^{-1}$ while for $\mathcal{R} \gg 1$ the dimensionless learning time scales approximately as $\tau \sim \mathcal{R}^{-1/2}$.

\subsection*{Simulation results}

To test the generality of these results, we turn to numerical simulations on a larger network with $N=64$ nodes and $N_e = 143$ edges and use nonequilibrium coupled learning (Eq.~\ref{eq:CoupledFlow}), to train a different allosteric task, involving $M_S=5$ source nodes, with applied voltages sampled from a normal distribution $V_S\sim\mathcal{N}(0,1)$, and $M_T=3$ target nodes with desired voltages sampled from $V_T\sim \mathcal{N}(0,0.2^2)$. We assess the success of learning with the equilibrium mean-squared error (MSE) function $C^*\equiv \sum_T(V_T-v_T^{*F})^2$, normalized by its initial value, and calculated as learning degrees of freedom are evolving. Note that this cost function is in equilibrium (once the physical degrees of freedom reached steady state), compared to the experimentally computed nonequilibrium cost function $C$ (for simulations of the nonequilibrium cost function see Methods).

Despite important differences between the experiment and simulations, the simulations capture all of the major features of the experiment (see Methods for details). In Fig. 2(a), error values $C^*$ during training are shown for different values of the scaled learning rate $\mathcal{R} \equiv \gamma_{\ell}/\gamma_p$, where $\gamma_{\ell}$ sets the learning rate and $\gamma_p$ sets the rate of physical relaxation as defined in Eq.~\ref{eq:CoupledFlow}. Here, we set the time scale with $\gamma_{\ell}$, so that a unit time corresponds to one training step. In the quasistatic regime where the physical dynamics are rapid compared to the learning degrees of freedom, $\mathcal{R} \ll 1$, the number of required training steps is independent of $\mathcal{R}$ but the real training time speeds up linearly with $\mathcal{R}$ (reddest curves in Fig. 2(a) and (b)). As in the experiment, $\mathcal{R}$ can be increased to $\mathcal{R} \approx 1$ with little effect on learning. Once $\mathcal{R}>\mathcal{O}(1)$, we observe error oscillates yet the network still learns successfully at longer times. The error oscillations strengthen and then weaken as the scale learning rate $\mathcal{R}$ increases further.

Fig. 2(c) shows similar results for more complex allosteric tasks with $10$ source nodes and $3$ target nodes trained on a network with $64$ nodes; results are averaged over $50$ different choices of such tasks. As seen experimentally and computationally for the simpler tasks in Fig. 2(a,b), the system learns with nearly equal low error for all $\mathcal{R} < \mathcal{O}(1)$.  Moreover, the achievable error can reach low values even when the learning degrees of freedom relax extremely rapidly relative to the physical degrees of freedom ($\mathcal{R} \gg 1$). In Fig. 2d, we plot the real time $\tau$ taken for the network to be trained to a normalized error threshold $C^*(\tau)=10^{-6}C^*(0)$. As in the experiment, we see a change of scaling in $\tau$ at $\mathcal{R} \sim 1$; for $\mathcal{R} \ll 1$ we see the expected linear improvement in the training time $\tau\sim \mathcal{R}^{-1}$. In the far-from-equilibrium regime, $\mathcal{R} \gg 1$, training becomes faster, but with slower scaling $\tau\sim \mathcal{R}^{-1/2}$.

\subsection*{Theoretical analysis}

To understand the observed behavior we take a second time derivative of the physical degrees of freedom in Eq.~\ref{eq:CoupledEquations} to obtain a damped harmonic oscillator equation:

\begin{equation}
\begin{aligned}
\gamma_p^{-2} \delta &\ddot{v}_a^{F} + \gamma_p^{-1}H^F\delta \dot{v}_a^F + \mathcal{R}(D^F)^2\delta v_a^F = 0
\end{aligned}
  \label{eq:Coupled2ndOrder}
\end{equation}

where we have defined the difference between the free and desired configurations $\delta v^F_T \equiv v^F_T - V_T$, and the physical Hessian and cross derivatives $H^F\equiv \partial_v^2 \mathcal{P}^F, D^F\equiv \partial_v \partial_G \mathcal{P}^F$. In addition to an inertial term there is an overdamped term for physical relaxation and a restoring one that scales with $\mathcal{R}$. As $H^F, D^F$ do not depend on the rates $\gamma_\ell$ and $\gamma_p$, the restoring term becomes important only when learning is fast enough ($\mathcal{R} \gtrsim 1$). The solutions to this equation become oscillatory when the discriminant $(H^F)^2-4\mathcal{R}(D^F)^2$ becomes negative at a value that we call $\mathcal{R}_c \approx 1$. For slow learning $\mathcal{R}\ll 1$, the decay time of the cost function $C\sim\delta v_a^2$ is dominated by the slowest decay mode of Eq.~\ref{eq:Coupled2ndOrder}, scaling as $\tau\sim \mathcal{R}^{-1}$. Beyond the critical learning rate $\mathcal{R}\gg 1$, it results from Eq.~\ref{eq:Coupled2ndOrder} that the oscillation frequency increases, scaling as $\mathcal{R}^{1/2}$. This frequency scaling dictates the training time $\tau\sim\mathcal{R}^{-1/2}$, which can be estimated by the first pass of the oscillatory dynamics through zero (see Methods for details).
These predictions are verified numerically in Fig. 3(a) for a network with $N=64$ nodes trained to a complex allosteric task with $M_S=10, M_T=3$. The oscillatory learning dynamics converge to solutions $\delta v_a \rightarrow 0$ as long as the physical derivatives $H^F,D^F$ vary slowly relative to the physical and learning degrees of freedom $v_a,G_j$, i.e. if the physical cost function (dissipated power) landscape is smooth enough with respect to the learning degrees of freedom. Rugged landscapes with narrow attractors would lead to physical dynamics that jump rapidly between basins, so that convergence cannot be guaranteed. The systems studied here have physical landscapes that are convex with a single basin so that they may be trained extremely fast.

These results also explain the suppression of learning oscillations. Expanding the contrastive function in series, we see that $\partial_G \mathcal{C} \approx -D^F \delta v_a^F$. If the learning dynamics pass through a flat region, where $\mathcal{C}$ changes slowly compared to the learning parameters $G_j$, the oscillatory term in Eq.~\ref{eq:Coupled2ndOrder} is suppressed, and the dynamics may become overdamped. To test this, we train a $N=64$ network for an allosteric task with $M_S=10,M_T=3$ at different $\mathcal{R}$. After training, we compute the Hessian of the contrastive function $\partial_G^2\mathcal{C}$ and its top eigenvalues. Smaller eigenvalues indicate flatter contrastive landscapes around the solution. In Fig. 3(b), we show that the landscape becomes flatter linearly with increasing $\mathcal{R}$. In these flow networks, the learned solutions become flatter faster than the increase in oscillation amplitude $\sim\mathcal{R}^{1/2}$, explaining the suppression of oscillations at high learning rates in Fig. 2(a,b).

\section*{Discussion}

We have demonstrated that physical learning does not need to be restricted to the quasistatic limit in which physical degrees of freedom equilibrate rapidly compared to the learning degrees of freedom. Indeed, the learning process can be sped up by many orders of magnitude by updating learning degrees of freedom at a rate that is comparable to the relaxation rate of the physical degrees of freedom without any qualitative change of behavior or much quantitative change in the achievable error. It is notable that the simulations recover the experimentally-observed phenomenology despite the differences between the simulated and experimental networks, suggesting that nonequilibrium coupled learning is robust to small alterations in the equations governing learning, as well as to noise and bias.

Our results may extend beyond contrastive learning in the physical learning electrical networks studied here. Similar results might arise in directed aging  ~\cite{pashine2019directed,hexner2019effect,hexner2019periodic} or contrastive learning~\cite{pashine2021local}, which so far have been carried out in the quasistatic limit in mechanical networks, where the physical degrees of freedom are the node positions, the physical cost function is the elastic energy and the learning degrees of freedom are edge stiffnesses or equilibrium lengths~\cite{pashine2019directed,hexner2019effect,hexner2019periodic} or the presence or absence of edges~\cite{pashine2021local}. The insight obtained here 
may shed light on certain forms of biological learning, where the plasticity time scales of some neural circuits are similar to the rate for reaching steady state~\cite{marom2010neural,sagi2012learning}. 

However, our results probably have their most direct application in machine learning in artificial neural networks. There, the quasistatic condition for recurrent networks constitutes a computational bottleneck for equilibrium propagation~\cite{bartunov2018assessing}. Our results show that quasistaticity is not mandatory, significantly reducing the computational time required for training recurrent networks, making local rules possible competitors with backpropagation algorithms.

Our results also show, however, that there are limits to how far the learning rate can be increased. We see that the learning rate can be sped up to be comparable to the physical rate of relaxation, $\mathcal{R} \sim 1$, without any penalty, and that the learning dynamics are overdamped up to that point. Beyond a critical rate $\mathcal{R}_c$ the learning dynamics develop under-damped oscillations, because the learning degrees of freedom evolve too rapidly and overshoot desired solutions. These oscillations become faster and stronger at higher learning rates but can be suppressed when the cost function landscape is sufficiently flat with respect to the learning degrees of freedom. This case is generic in the over-parameterized regime of deep learning. In complex non-convex landscapes underlying learning problems in the under-parameterized regime, we expect fast training to fail as the network rapidly jumps between basins in the physical cost function landscape due to the changing learning parameters. In the flow networks we trained for relatively few allosteric tasks, which lie in the convex, over-parameterized regime, we observe that training is possible and successful far from equilibrium.

We trained physical networks for allosteric tasks, where a network learns a mapping between a single set of inputs and a single set of outputs. More complex learning problems such as classification, where networks are trained with multiple input and output examples, require further analysis. Training a network for multiple tasks introduces a new timescale corresponding to the rate at which the system is fed training examples, $\sigma$. The insight from our results here suggests that training for multiple tasks may only succeed if the task switching rate is lower than the rate of physical relaxation $\sigma \lesssim \gamma_p$. Otherwise, the system will not effectively evolve under the influence of the training example. We leave a more detailed account of the effect of task switching on far-from-equilibrium learning for future study.

\section*{Materials and Methods}
%

\subsection*{Nonequilibrium learning in physical resistor networks}

In this work we train a physical learning network~\cite{dillavou2021demonstration}, a self-adjusting system of variable resistors, to perform allosteric tasks. In such a task, the system succeeds if, given a set of applied input voltage(s), the desired output voltage(s) are generated (as a physical response) at the output node(s). Error is measured as the mean-squared deviation from those desired voltages. Our system consists of two identical networks of variable resistors, the `free' network that generates outputs given imposed inputs, and the `clamped' network whose outputs are also imposed at a value preferable to the free state output (closer to the goal.) Internal circuitry compares the free and clamped network voltage drops on each edge, and determines whether to increase or decrease the resistance of that edge. Resistances are updated as
\begin{equation}
    \Delta R_j =  
     \begin{cases}
      +\delta R & \text{if} \ |\Delta v^C_j| > |\Delta V^F_j|, \\
      -\delta R & \text{otherwise.}
    \end{cases}  
    \label{sign}
\end{equation}
This update rule is feasible for our resistive elements, digital potentiometers with 128 possible resistance values, each separated by $\delta R \sim 781 \Omega$. This update rule is implemented using two comparators and an XOR gate, specifically
\begin{equation}
     \Delta R_j =   
     \begin{cases}
      +\delta R & \text{if} \ \textrm{XOR}\left[\Delta v_j^C > \Delta v_j^F, 0 < \Delta v_j^C\right] \\
      -\delta R & \text{otherwise}
    \end{cases}  
    \label{simplerule}
\end{equation}
Eq \ref{sign} and \ref{simplerule} are identical provided the voltage drop across each network is the same sign, which is nearly always the case in our system. This rule is subject to experimental noise in the comparator readings, of order $0.01V$, which generates a noise floor, seen in Fig. 1(a) and (b).

The `supervisor' (CPU) is responsible only for applying the input voltages to both networks, measuring the free state outputs ($v_T^F$), and applying the clamped state outputs ($v_T^C$) closer to the goal ($V_T$) according to 
\begin{equation}
    v_T^C(t)=v_T^F(t)+\eta (V_T - v_T^F(t))
\end{equation}
In the experimental system, $\eta \ll 1$  leads to measurement errors and poorer performance, and thus $\eta = 0.5$ was used. Edge resistance changes are triggered by a global clock signal. 

The system (or rather the physics within the system) perform the `computation' of the output voltages naturally. In previous work~\cite{dillavou2021demonstration} these responses were extremely rapid, but here we slow these physical dynamics down by adding capacitors in parallel to each edge of the network. In particular, we modulate the physical relaxation rate by changing the value of these capacitors. After applying new clamped voltages or after resistance values are updated, we allow the voltages to relax for time $1/\gamma_{\ell}$. In particular, we modulate the learning rate by adjusting the length of this pause. In the experimental system, all errors reported are \textit{out of equilibrium}, that is, these are the deviations of the voltages as measured during the experiment from the goal values.

\subsection*{Simulating nonequilibrium learning}

We simulate variable resistor networks that are trained to perform allosteric tasks, using an approach called coupled learning~\cite{stern2021supervised}. An allosteric task involves clamping some source nodes with prescribed voltages $V_S$. The network is trained to produce target voltages at certain other nodes $V_T$. Nonequilibrium training is performed by time integration of the dynamics given by Eq.~3. We interpret these dynamics as applied to two networks whose conductance values $G_j(t)$ evolve together. The first is the free network (Fig. 4(a)), with voltage values $v_a^F(t)$, for which the source nodes are always held at $V_S$. In addition, a clamped network (Fig. 4(b)) with voltages $v_a^C(t)$ is constrained such that the source nodes are again held at $V_S$. The target constraints are supplied by a supervisor, setting the target voltages at $v_T^C(t)=v_T^F(t)+\eta (V_T - v_T^F(t))$, with a small nudge parameter $\eta=10^{-3}$.


Resistor networks naturally minimize their power dissipation $\mathcal{P}^{F,C}=\sum_j G_j (\Delta v_j^{F,C})^2/2$ subject to external constraints, where $\Delta v_j = \Delta_{ja} v_a$ are the voltage drops across edges, and $\Delta_{ja}$ is the incidence matrix, describing the directed connections between nodes. Therefore, the (overdamped) dynamics of the free and clamped network without learning is given by $$\dot{v}_a = -\gamma_p \partial_v \mathcal{P}= -\gamma_p \sum_j \Delta_{aj}^T G_j \Delta_{ja} v_a$$ 
When learning is turned on $\gamma_{\ell} > 0$, the coupled learning rule (Fig. 4(c)) describes the dynamics of the edge conductances $G_j$ in terms of the contrastive function, or the difference between the free and clamped states $\mathcal{C}\equiv \eta^{-1}[P^C-P^F]$, $$\dot{G}_j(t) = -\gamma_\ell\partial_G\mathcal{C}=\frac{1}{2}\gamma_\ell\eta^{-1}[(\Delta v_j^F(t))^2-(\Delta v_j^C(t))^2]$$

These two sets of equation are numerically integrated for resistor networks using the Runge-Kutta method implemented in SciPy.integrate. To set the initial condition, we initialize all conductance values uniformly at $G_j(t=0)=1$. We then apply the source (and target) constraints on the free and clamped networks, allowing them to first reach a steady state with $\gamma_p=1, \gamma_\ell=0$. Nonequilibrium training is then commenced by setting the learning rate $\gamma_\ell$ to achieve a scaled learning rate of interest $\mathcal{R}$.

Training success is measured by the equilibrium value of the cost function $C^*(G_j)\sim [V_T-v_T^{*F}(G_j)]^2$. This is done by setting the conductance values $G_j$ by their values at time $t$ during training, and then computing the steady state free target voltages $v_T^{*F}$. The nonequilibrium cost function is easier to compute $C(t)\sim [V_T-v_T^{F}(t)]^2$. In the main text, we generally normalize the cost function by its value at $t=0$, so that the constant pre-factor is unimportant. This also allows the comparison of different tasks, which all initially have different cost function values.

While the simulations mimic physical dynamics and learning in electric resistor networks, there are several notable differences between them and the experiment. Firstly, the physical dynamics of the simulation are given by overdamped minimization of the dissipated power $\dot{v}_a=-\gamma_p \partial_v \mathcal{P}$. This is similar to assuming a `global' capacitance for the network, rather than separate capacitors connected in parallel to each resistor, as done in the experiment. Perhaps more importantly, the learning rules used are significantly different: The simulations used the original coupled learning rule $\dot{G}_j=-\gamma_\ell\partial_G\mathcal{C}$, while the experiment uses a signed version of this learning rule $\Delta G_j =\Delta R_j^{-1}= -g^{-2}_j \delta R\cdot \mathrm{sgn}(\partial_G\mathcal{C})$ due to the discrete nature of the experimental variable resistors. Finally, while the errors measure in the experiment where (instantaneous) nonequilibrium errors $C(t)$, in Fig.~2 we showed simulation results for the associated equilibrium errors $C^*(t)$, measured after the voltages are allowed to equilibrate.

While the physical dynamics are different between the experiment and simulation, we note that both types equilibrate to the same state, minimizing the dissipated power $\mathcal{P}$. We therefore do not expect much difference between them for slow learning $\mathcal{R}\ll 1$. Even far from equilibrium, while $\dot{v}_a$ might not be the same for the two types of dynamics, successful learning will eventually bring both systems to physical steady states. The difference in learning rules between the two systems was discussed extensively in Ref.~\cite{dillavou2021demonstration}. Most notably, the experimental signed learning rule tends to reach simple limit cycles, where the system jumps between two states of the resistance vectors, corresponding to two alternating error values. This property, together with the discreteness of the experimental resistances and other sources of noise, gives rise to a finite error floor in the experiment, absent from the simulations.

We have opted to mainly discuss equilibrium error in simulations, as the equilibrium error $C^*(t)$ is more appropriate for assessing the functionality of the system (as it is history independent). However, we have also measured the nonequilibrium errors $C(t)$ in the simulation, observing similar results to those discussed for the equilibrium errors. In Fig. 5, we show the nonequilibrium errors in simulations for different values of the scaled learning rate $\mathcal{R}$. In contrast with the equilibrium errors, and similarly to the experimental results, these errors oscillate but never overshoot the initial errors $C^*(t=0)=1$. Other than that, there are no qualitative differences between the equilibrium $C$ and the nonequilibrium $C^*$, both exhibiting under-damped learning oscillations at $\mathcal{R}\gtrsim 1$.

\subsection*{Nonequilibrium learning dynamics}

The physical overdamped dynamics of the free and clamped states are given by:

\begin{equation}
\begin{aligned}
 \dot{v}_a^{F,C} &= -\gamma_p \partial_v \mathcal{P}^{F,C}(v_a^{F,C};G_j)
 \label{eq:E1}
 \end{aligned}
\end{equation}
 
where $v^F_a$ are the free physical degrees of freedom, $v^C_a$ are the clamped physical degrees of freedom, and $G_j$ are the learning degrees of freedom. $\mathcal{P}^F$ and $\mathcal{P}^C$ are the physical cost function (e.g. energy, power) of the free state and clamped states, and $\gamma_p$ the rate of the physical dynamics. Recall that we define the clamped state by nudging the physical degrees of freedom toward their desired values $v^C_T = v^F_T + \eta(V_T-v_T^F)$, with $\eta \ll 1$ the nudge parameter. 

Given a small nudge $\eta \ll 1$, we perform a Taylor expansion for the clamped physical cost function around its free counterpart, assuming for simplicity that all physical degrees of freedom may be nudged:

\begin{equation}
\begin{split}
\mathcal{P}^C(v_a^C;G_j)&=\mathcal{P}^C(v_a^F+\eta(V_a-v_a^F);G_j) \\
& \approx \mathcal{P}^F + \eta(V_a-v_a^F)^T \partial_v \mathcal{P}^F + \frac{1}{2}\eta^2 (V_a-v_a^F)^T\partial_v^2 \mathcal{P}^F (V_a-v_a^F) + \dots\\
& \approx \mathcal{P}^F + \eta(V_a-v^F_a)^T \partial_v \mathcal{P}^F + \frac{1}{2}\eta^2 (V_a-v_a^F)^T H^F (V_a-v_a^F),
\end{split}
\label{eq:E4}
\end{equation}

where we defined the free Hessian $H^F\equiv\partial^2_v \mathcal{P}^F(v_a^F;G_j)$ as the second derivative of the free physical cost function with respect to the physical degrees of freedom. In the following we set $v^F_a\equiv v_a$. Note that when the free dynamics are equilibrated (i.e. $\partial_v \mathcal{P}^F = 0$), the linear term in $\eta$ vanishes. Now we define the contrastive function $\mathcal{C}$ as the difference of the clamped and free physical cost functions.  

\begin{equation}
\begin{split}
 \mathcal{C} &\equiv \eta^{-1} [\mathcal{P}^C - \mathcal{P}^F] \approx (V_a-v_a)^T \partial_v \mathcal{P}^F + \frac{1}{2}\eta (V_a-v_a)^T H^F (V_a-v_a) 
\end{split}
\label{eq:E5}
\end{equation}

Note that in general it is different than the cost function $C$, the difference between the desired and observed outputs. When the learning degrees of freedom are equilibrated $\partial_v {P}^F=0$ only the second term exists, which is positive semi-definite. However, out of equilibrium, the first term dominates and the contrastive function may become negative (meaning it is \emph{not} a proper cost function). The dynamics of the learning degrees of freedom are given by gradient descent on the contrastive function, with learning rate $\gamma_\ell$.

\begin{equation}
\begin{split}
 \dot{G}_j &= -\gamma_\ell \partial_G \mathcal{C} = -\gamma_\ell (V_a-v_a)^T \partial_G\partial_v \mathcal{P}^F - \frac{1}{2}\gamma_\ell \eta (V_a-v_a)^T \partial_w H^F (V_a-v_a)  \\
 &\approx \gamma_\ell (\partial_G\partial_v \mathcal{P}^F) (V_a-v_a) \equiv \gamma_\ell D^F (V_a-v_a)
 \label{eq:E6}.
 \end{split}
\end{equation}

Since we are focused on nonequilibrium learning dynamics, we kept only the first dominating term and further defined the cross derivative of the free energy $D^F\equiv \partial_G\partial_v \mathcal{P}^F$.

Performing a second time derivative of the physical degrees of freedom:

\begin{equation}
\begin{split}
 \ddot{v}_a &= -\gamma_p \frac{d}{dt} \partial_v \mathcal{P}^F = -\gamma_p \partial_v \frac{d}{dt}\mathcal{P}^F =  -\gamma_p  \partial_v [\dot{v}_a\partial_v \mathcal{P}^F + \dot{G}_j\partial_G \mathcal{P}^F]\\
 &= -\gamma_p [\dot{v}_a\partial^2_v \mathcal{P}^F + \dot{G}_j\partial_v\partial_G \mathcal{P}^F]= -\gamma_p[H^F \dot{v}_a+ \gamma_\ell (D^F)^2 (V_a-v_a)]\\
 \ddot{v}_a& + \gamma_p H^F \dot{v}_a + \gamma_p \gamma_\ell (D^F)^2 (V_a-v_a) =0
 \label{eq:E7}.
 \end{split}
\end{equation}

Let us further define $\delta v_a\equiv v_a-V_a$ and divide the previous equation by $\gamma^2_p$:

\begin{equation}
\begin{split}
 &\gamma^{-2}_p\delta\ddot{v}_a +  \gamma^{-1} H^F\delta\dot{v}_a  + \mathcal{R} (D^F)^2 \delta v_a = 0\\
 &\gamma^{-1}_p \dot{G}_j \approx \mathcal{R} D^F \delta v_a 
 \label{eq:E8}
 \end{split}
\end{equation}

This last set of equations shows that for slow learning $\mathcal{R} \ll 1$, the dynamics of the physical degrees of freedom $\delta v_a$ are overdamped. However, when learning becomes faster, the oscillatory term becomes more important and both physical and learning degrees of freedom will start exhibiting under-damped oscillations at a critical rate $\mathcal{R}_c$. 

Suppose that the derivatives $H^F, D^F$ change slowly with $\delta v_a, G_j$. In that case, Eq.~\ref{eq:E8} describes a damped harmonic oscillator with a solution to the dynamics of the physical degrees of freedom $\delta v_a$ approximately given by:

\begin{equation}
\begin{split}
\delta v_a(t) & \approx A_{a,+} e^{-\omega_+ t} + A_{a,-}- e^{-\omega_- t} \\
\omega_{a,\pm} & = \frac{1}{2} \gamma_p (H^F \pm \sqrt{(H^F)^2-4\mathcal{R}(D^F)^2}),
 \label{eq:FormalSol}
 \end{split}
\end{equation}

Where $A_{a,\pm}$ are amplitudes for the two families of eigenmodes, determined by initial conditions. $\omega_{a,\pm}$ are the frequencies of the dynamics, and should be interpreted as the eigenvalues of the matrices $\gamma_p (H^F \pm \sqrt{(H^F)^2-4\mathcal{R}(D^F)^2})$.  When learning is slow $\mathcal{R}\ll 1$, all modes $\omega$ have real positive frequencies, so that $\delta v_a$ decays to zero with a rate determined by the slowest frequency in $\omega_{a,-}$. In this case we can approximate the slowest mode by $\omega_-\approx \gamma_\ell \times (H^F)^{-1}(D^F)^2$. Noting that the cost function is $C\sim \delta v_a^2$, explaining why close to equilibrium the effective training time decays like $\tau \sim \omega^{-1}_- \sim \mathcal{R}^{-1}$.

When learning is sped up, a critical point is reached for which there is a mode with $(H^F)^2-4\mathcal{R}_c(D^F)^2=0$. Then the square root becomes imaginary, and we see oscillation in $\delta v_a$. When the discriminant becomes negative for all modes, the real parts of their frequencies $\mathrm{Re}[\omega_a]$ are given by the eigenvalues of $\gamma_p H^F/2$. For fast learning far from equilibrium ($\mathcal{R}\gg 1$), the oscillatory parts of the modes $\mathrm{Im}[\omega_a]$ are given by eigenvalues of $\gamma_p\sqrt{\mathcal{R} }D^F/2$, so that oscillations become faster as the square root of the scaled rate $\mathcal{R}^{1/2}$. We can further ask how much time it takes for the $\delta v_a$ to vanish for the first time. For $\mathcal{R}\gg 1$ we may approximate the cost function as $C\sim \delta v_a^2\sim \cos ^2(\gamma_p\sqrt{\mathcal{R}}D^F t)\exp(-\gamma _p H^F t)$. This cost function first vanishes when $\gamma_p\sqrt{\mathcal{R}}D^F \tau = \pi/2 \rightarrow \tau \sim \mathcal{R}^{-1/2}$, suggesting the scaling found in the experiments and simulations.


\bibliography{sciadvbib}
\bibliographystyle{ScienceAdvances}

\noindent \textbf{Acknowledgements:} 
%
We thank Sean Fancher, Eleni Katifori and Vijay Balasubramanian for insightful discussions.\\
\noindent \textbf{Funding:} This work was supported by the U.S. Department of Energy, Office of Basic Energy Sciences, Division of Materials Sciences and Engineering award DE-SC0020963 (M.S.), and the National Science Foundation via the UPenn MRSEC/DMR-1720530 (S.D. and D.J.D.), DMR-2005749 (A.J.L.) and Army Research Office award W911NF-21-1-0076 and air force office of scientific research award (FA9550-21-1-0313) (M.Z.M.).\\
\noindent \textbf{Author Contributions} All authors conceived the research. MS designed and conducted the simulations and analyses. SD designed and conducted the experiments. All authors wrote the manuscript.\\
\noindent \textbf{Competing Interests} The authors declare that they have no competing financial interests.\\
\noindent \textbf{Data and materials availability:} Additional data and materials are available upon request.


\clearpage

\begin{figure}[h!]
\includegraphics[width=0.70\linewidth]{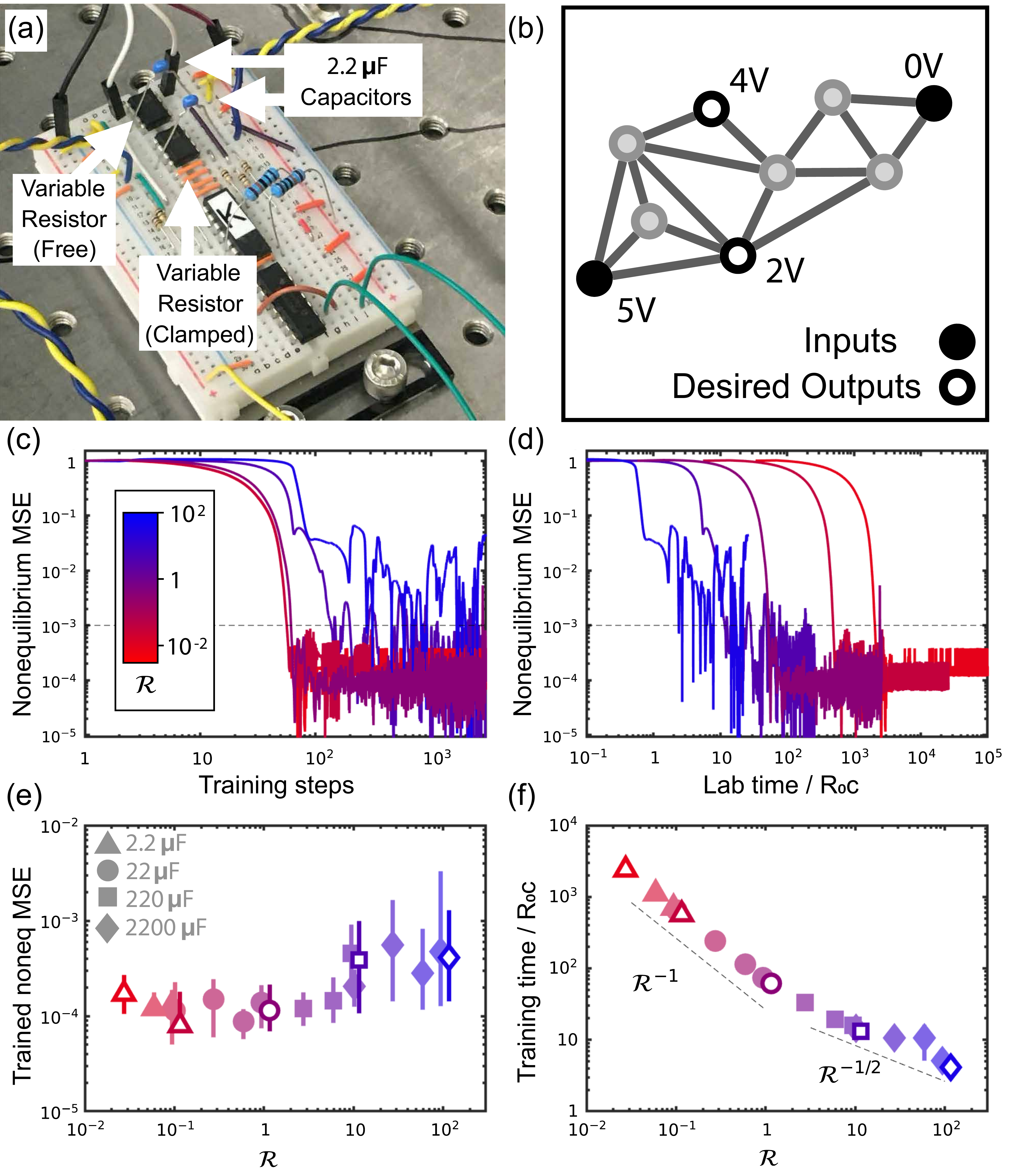}
\label{fig:Exp}
\end{figure}
\noindent {\bf Fig. 1.} Nonequilibrium learning in a physical learning network. a) A single edge (in both free and clamped networks) with parallel 2.2$\mu$F capacitors 
highlighted with arrows. b) Network structure with inputs and outputs for the allosteric task used in (c)-(f). c) Typical nonequilibrium (instantaneous) mean squared error (MSE) traces, divided by initial error value, for an allosteric task as a function of training steps and d) real lab time. Colors indicate the scaled learning rate $\mathcal{R}$, and dotted line shows error threshold for (f). e) Nonequilibrium MSE for networks trained for $3\times 10^3$ steps as a function of $\mathcal{R}$. Error bars indicate first and third quartiles. Shapes indicate capacitor values used, with hollow points corresponding to the traces shown in (c) and (d). f) Average training times $\tau$ when the system achieves a nonequilibrium MSE below a threshold of $\mathrm{MSE}(\tau)=10^{-3}\mathrm{MSE}(0)$. 
Dotted lines are power laws of $-1$ and $-1/2$ respectively.

\begin{figure}[h!]
\includegraphics[width=0.70\linewidth]{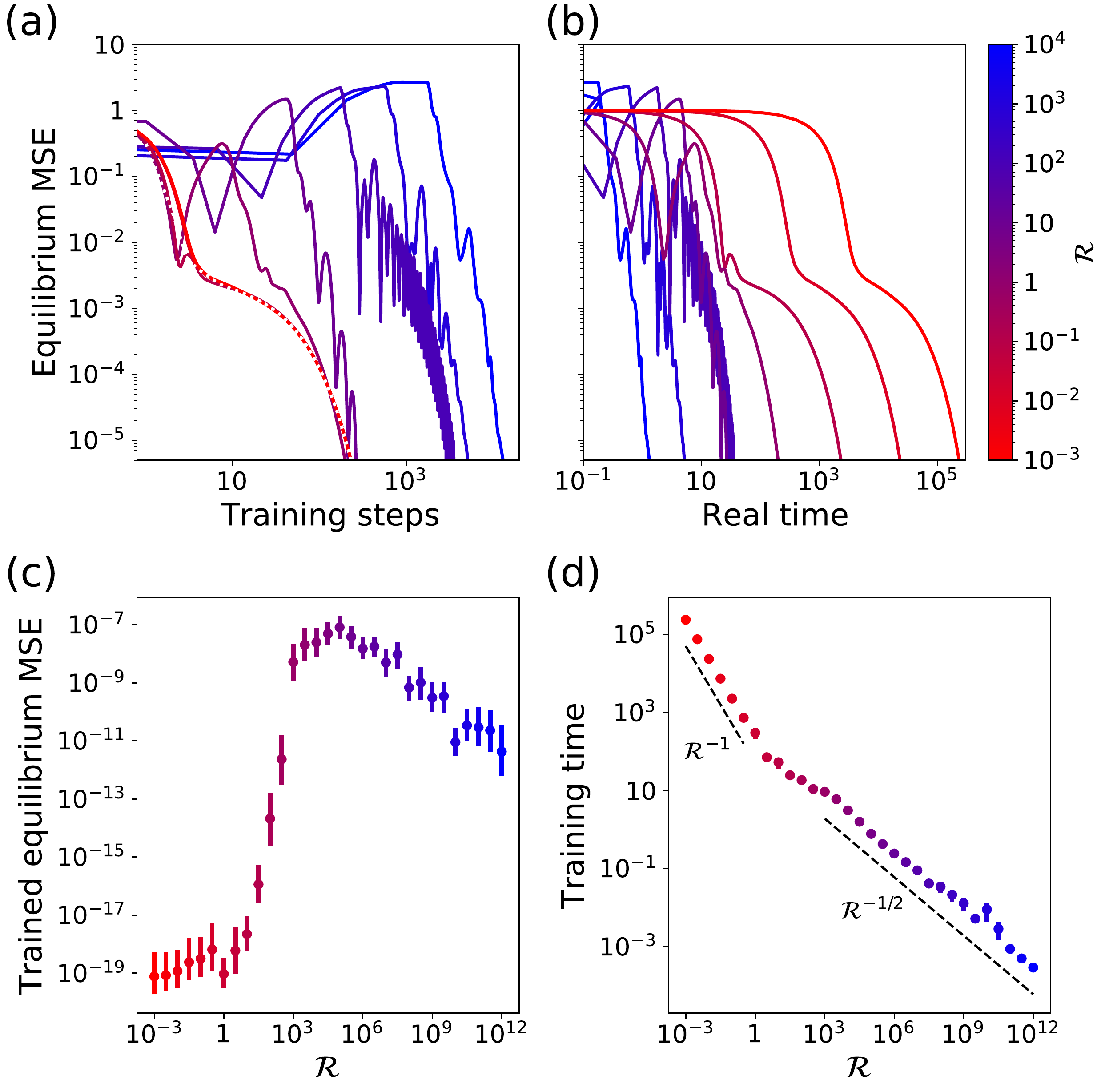}
\label{fig:Sim}
\end{figure}
\noindent {\bf Fig. 2.} Nonequilibrium in silico training of resistor networks. a) Network of size $N=64$ is trained for an allosteric task with $M_S=5$ sources and $M_T=3$ targets. We plot the equilibrium mean squared error (MSE) during training at different scaled learning rates $\mathcal{R}$. When $\mathcal{R} \gtrsim 1$, more training steps are required to reduce the error (dotted line denotes the quasistatic limit $\mathcal{R} \rightarrow 0$). b) However, training with a higher learning rate allows the system to learn more rapidly in real time. c) The average MSE of trained networks ($N=64, M_S=10, M_T=3$) shows that comparable success is achieved  for learning rates up to $\mathcal{R} \approx 1$. While not as accurate, learning still substantially reduces errors even far from equilibrium, where the learning degrees of freedom relax rapidly relative to the physical ones $\mathcal{R} \gg 1$. d) 
Training time $\tau$ until the equilibrium MSE reaches a threshold $\mathrm{MSE}^*(\tau)=10^{-6}\mathrm{MSE}^*(0)$. As in the experiment, training time shortens linearly like $\mathcal{R}^{-1}$ for slow learning $\mathcal{R} \ll 1$, while for fast learning $\mathcal{R} \gg 1$ training times shrink as $\mathcal{R}^{-1/2}$.

\begin{figure}[h!]
\includegraphics[width=0.70\linewidth]{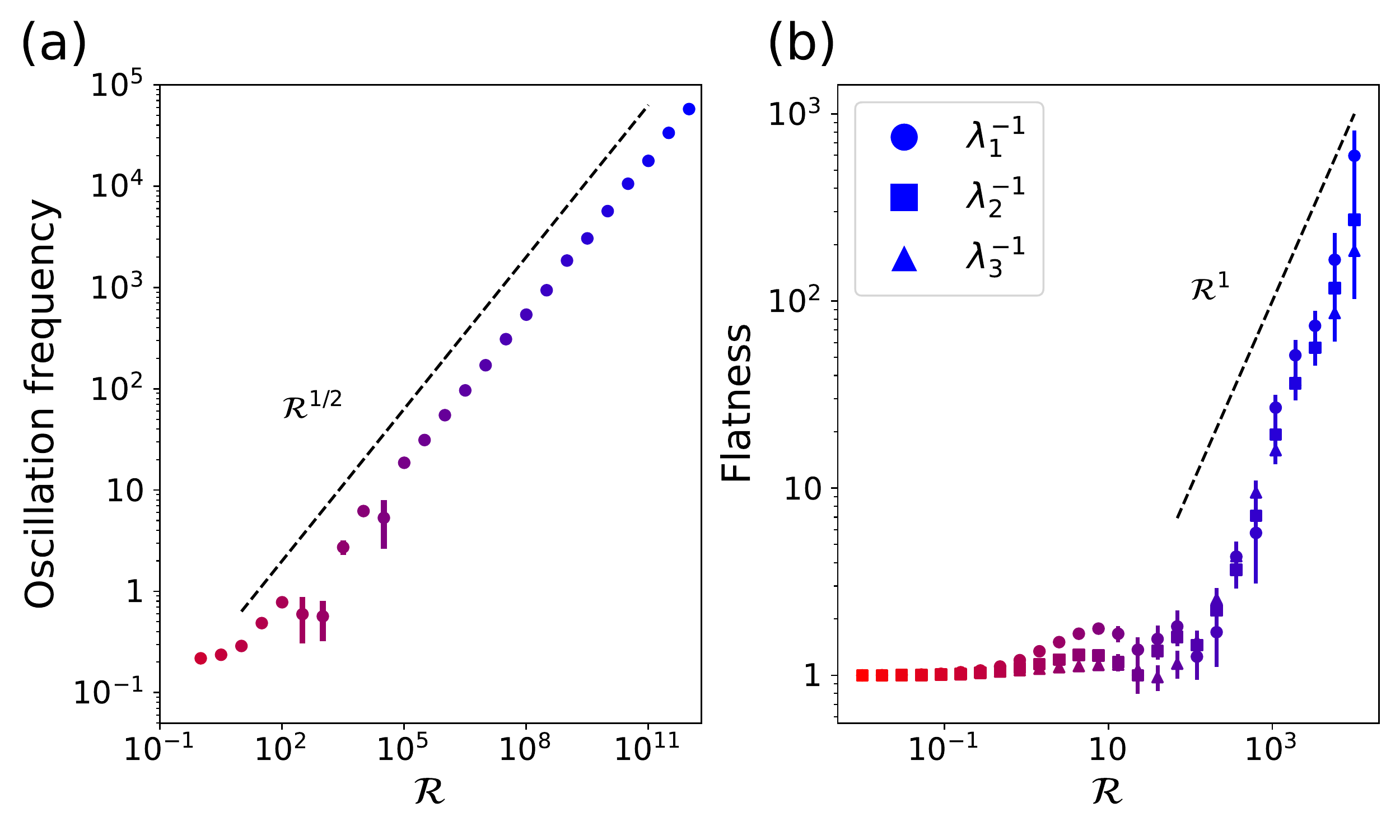}
\label{fig:Osc}
\end{figure}
\noindent {\bf Fig. 3.} Nonequilibrium dynamical effects for fast training of allosteric tasks in resistor networks. a) Starting at a critical learning rate $\mathcal{R}_c\approx 1$, the training error exhibits oscillations whose frequency grows as $\mathcal{R}^{1/2}$. b) Learning oscillations are suppressed as at higher learning rates, the network finds solutions of flatter cost, as indicated by the diminishing lead eigenvalues of the contrastive Hessian $\partial_G^2 \mathcal{C}$.

\begin{figure}[h!]
\includegraphics[width=0.7\linewidth]{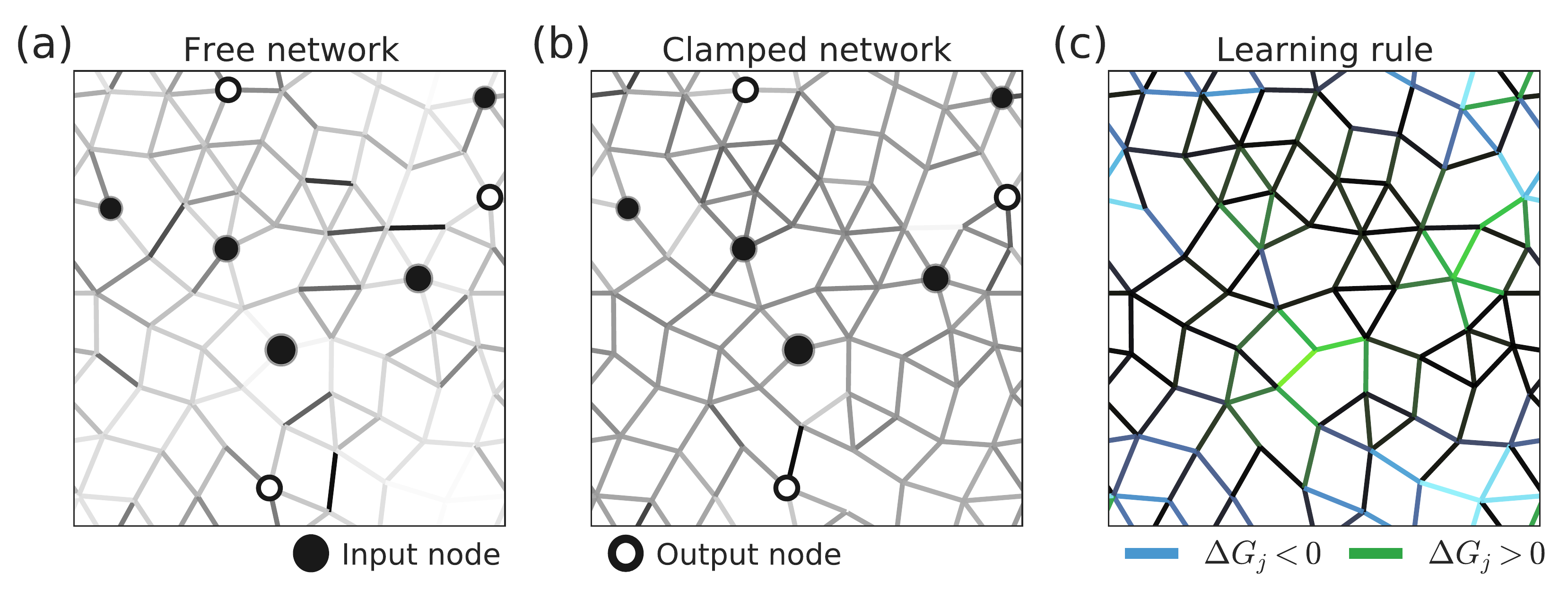}
\label{fig:Net}
\end{figure}
\noindent {\bf Fig. 4.} Coupled learning in electric resistor networks. a) In the free network, the input\textbackslash source nodes (full black circles) are held at fixed voltages $V_S$, and the output\textbackslash target node voltages $v_T^F$ (hollow circles) evolve by the network dynamics. b) In the clamped network, the source nodes are held at $V_S$ (similar to the free network), but the output node voltages are nudged to $v_T^C=v_T^F+\eta(V_T-v_T^F)$. c) At every edge, resistance value is modified in proportion to the difference in voltage drops (squared) between the free and clamped network $\Delta G_j\sim (\Delta v_j^F)^2 - (\Delta v_j^C)^2$.

\begin{figure}[h!]
\includegraphics[width=0.7\linewidth]{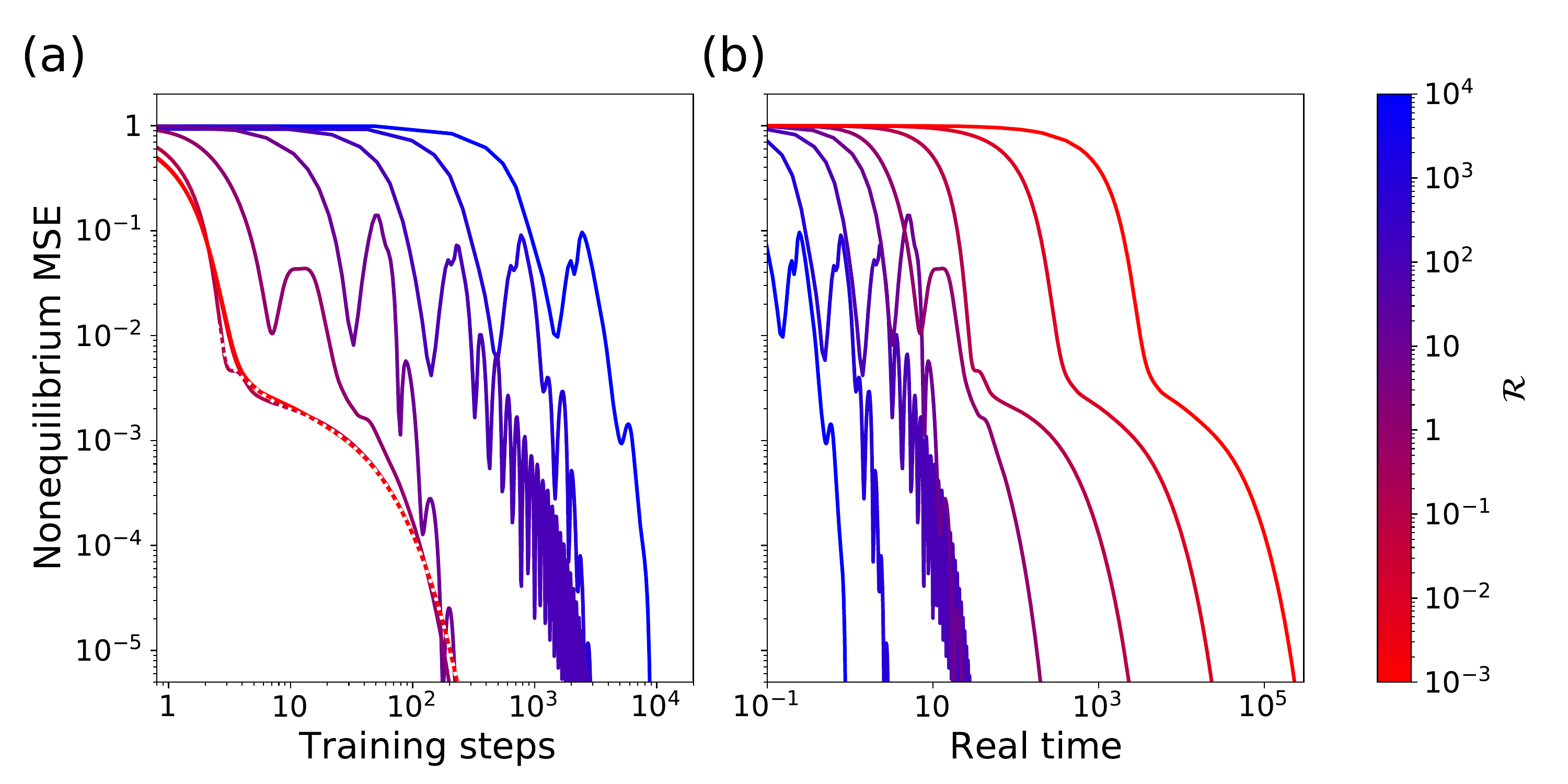}
\label{fig:NonEq}
\end{figure}
\noindent {\bf Fig. 5.} Nonequilibrium errors in learning simulations for a size $N=64$ network with $M_S=5$ sources and $M_T=3$ targets. a) Errors during training in terms of training steps and b) training time. The simulation results for nonequilibrium errors are qualitatively similar to those shown in the experiment, and also to the simulation equilibrium errors, except that the nonequilibrium errors generally do not overshoot their initial values.

\end{document}